\documentclass[10pt]{article}
\usepackage{graphicx}
\usepackage{amsmath, amssymb}
\usepackage{subfig}

\def\Title#1{\begin{center} {\Large #1 } \end{center}}
\def\Author#1{\begin{center}{ \sc #1} \end{center}}
\def\Address#1{\begin{center}{ \it #1} \end{center}}

\newcommand\pubblock{\rightline{\begin{tabular}{l} Proceedings of the Second Annual LHCP\\ \pubnumber\\
         \pubdate  \end{tabular}}}

\newenvironment{Abstract}{\begin{quotation} \begin{center} 
             \large ABSTRACT \end{center}\bigskip 
      \begin{center}\begin{large}}{\end{large}\end{center} \end{quotation}}

\newenvironment{Presented}{\begin{quotation} \begin{center} 
             PRESENTED AT\end{center}\bigskip 
      \begin{center}\begin{large}}{\end{large}\end{center} \end{quotation}}

\def\beq{\begin{equation}}
\def\eeq#1{\label{#1}\end{equation}}
\def\eeqn{\end{equation}}

\def\beqa{\begin{eqnarray}}
\def\eeqa#1{\label{#1}\end{eqnarray}}
\def\eeqan{\end{eqnarray}}

\let\bar=\overbar

\def\Dslash{\not{\hbox{\kern-4pt $D$}}}
\def\dslash{\not{\hbox{\kern-2pt $\del$}}}

\def\ee{e^+e^-}

\def\msb{{\bar{\ssstyle M \kern -1pt S}}}

\def\Hgg{\ensuremath{H \rightarrow \gamma\gamma}}
\def\H4l{\ensuremath{H \rightarrow ZZ^{*} \rightarrow 4\ell}}
\def\HWW{\ensuremath{H \rightarrow WW^{*} \rightarrow \ell\nu\ell\nu}}
\def\HZg{\ensuremath{H \rightarrow Z\gamma}}

\def\ee{\ensuremath{e^+e^-}}
\def\Zee{\ensuremath{Z \rightarrow e^+e^-}}
\def\Zmumu{\ensuremath{Z \rightarrow \mu^+\mu^-}}

\def\Jpsimumu{\ensuremath{J/\psi \rightarrow \mu^+\mu^-}}
\def\Upsmumu{\ensuremath{\Upsilon \rightarrow \mu^+\mu^-}}
\def\ttbar{\ensuremath{t\bar{t}}}
\def\ZZ4l{\ensuremath{ZZ^{*} \rightarrow 4\ell}}
\def\Zll{\ensuremath{Z \rightarrow \ell\ell}}

\def\pt{\ensuremath{p_\mathrm{T}}}
\def\mT{\ensuremath{m_\mathrm{T}}}
\def\ptt{\ensuremath{p_{\mathrm{Tt}}}}
\def\mgg{\ensuremath{m_{\gamma\gamma}}}
\def\m4l{\ensuremath{m_{4\ell}}}
\def\mH{\ensuremath{m_{\mathrm{H}}}}
\def\mllg{\ensuremath{m_{\ell\ell\gamma}}}

\def\ifb{fb$^{-1}$}
\def\X0{\ensuremath{X_0}}

\usepackage{color}

\newcommand\pubnumber{ ATL-PHYS-PROC-2014-135 }

\newcommand\pubdate{\today}

\def\affiliation{
On behalf of the ATLAS Collaboration, \\
CERN, departement PH \\
1211 Geneva 23, Switzerland }

\begin{document}

\large
\begin{titlepage}
\pubblock

\vfill
\Title{ Measurement of properties of the Higgs boson in bosonic decay channels using the ATLAS detector }
\vfill

\Author{ Bruno Lenzi }
\Address{\affiliation}
\vfill
\begin{Abstract}

The properties of the Higgs boson measured in bosonic decay channels \newline (\Hgg, \H4l, \HWW, \HZg) with 25~\ifb\ of $pp$ collision data from the Large Hadron Collider run-1 collected by the ATLAS experiment are presented. The results include an improved measurement of the mass of the Higgs boson from a combined fit to the invariant mass spectra of the decay channels \Hgg\ and \H4l, which yields \newline $\mH = 125.36 \pm 0.37 \mbox{ (stat)} \pm 0.18 \mbox{ (syst)} \mbox{ GeV}$ $= 125.36 \pm 0.41$~GeV \newline and supersedes the previous result from ATLAS.

\end{Abstract}
\vfill

\begin{Presented}
The Second Annual Conference\\
 on Large Hadron Collider Physics \\
Columbia University, New York, U.S.A \\ 
June 2-7, 2014
\end{Presented}
\vfill
\end{titlepage}
\def\thefootnote{\fnsymbol{footnote}}
\setcounter{footnote}{0}
%

\normalsize 


\section{Introduction}

\ \ \ \ The discovery of the Higgs boson announced by the ATLAS~\cite{Aad:2012tfa} and CMS~\cite{Chatrchyan:2012ufa} collaborations in July 2012 was mostly based on the results from bosonic decay channels \Hgg, \H4l and \HWW, $\ell = e, \mu$. The attention of the particle physics community is now focused on the measurement of its properties -- mass, production modes and couplings, spin and CP numbers -- to which these channels can largely contribute. On the other hand, the observation of the \HZg\ signal will be a major objective of both collaborations during the coming years of operation of the Large Hadron Collider (LHC). The present paper describes the measurement of the Higgs boson properties in these channels, obtained from the analysis of 25~\ifb\ of LHC $pp$ collision data recorded at center-of-mass energies of 7~TeV and 8~TeV by the ATLAS experiment~\cite{ATLAS_detector_paper}. An improved measurement of the mass of the Higgs boson from the decay channels \Hgg\ and \H4l, described in Ref.~\cite{ATLAS_Higgs_mass} and based on improved energy-scale calibrations for photons, electrons and muons, as well as other analysis improvements, is reported.

\section{Energy-scale calibrations of electrons, photons and muons}

\ \ \ \ The calibration strategy for the energy measurement of electrons and photons is described in detail in Ref.~\cite{ATLAS_egammaCalibPaper_run1} and proceeds in three main steps described below.

\begin{itemize}
\item A set of pre-corrections are applied to the data, in order to equalise the response from each cell of the electromagnetic (EM) calorimeter. The inter-calibration between the first two longitudinal layers of the EM calorimeter is obtained using muons from $Z$ boson decays in bins of the pseudorapidity $\eta$. The relative calibration of the presampler, which corrects for energy losses upstream the calorimeter, is extracted by comparing the energy response as a function of the longitudinal shower development in data and simulation. The stability of the energy response is found to be at the level of 0.05\% as a function of time and pileup (additional interactions in the same or neighbouring bunch crossings).

\item A multivariate regression algorithm, optimised on simulated data, is applied to correct the energy of EM objects, with a 10\% improvement in the expected mass resolution of the \Hgg\ decay compared to the previous results~\cite{ATLAS_Higgs_diboson}. The energies measured in the presampler and other calorimeter layers, the longitudinal shower development and the position of the cluster centroid are used as inputs. For photons that convert into \ee\ pairs, the track transverse momenta (\pt) and the conversion radius are also used.

\item A global energy scale adjustment and an effective constant term for the energy resolution are derived as a function of $\eta$ using \Zee\ decays. Corrections of typically 1\% -- 3\% to the energy scale and constant terms varying from 0.7\% for $|\eta| < 0.6$ up to 3.5\% in the transition region between barrel and endcap calorimeters are obtained by comparing the reconstructed mass in data and simulation. Good agreement between data and simulation is achieved after the corrections.

\end{itemize}

The determination of the muon momentum scale and resolution are described in detail in Ref.~\cite{ATLAS_muonPaper_run1} and studied with about 6 million \Jpsimumu\ and about 9 million \Zmumu\ decays from collision data. The reconstructed muon momentum in the simulation is corrected to match the observed invariant mass distributions of the resonances. Independent corrections are derived to the momenta measured in the Inner Detector (ID) and Muon Spectrometer (MS) in bins of $\eta$ and \pt, and following the MS sector granularity in $\phi$. The corrections to ID (MS) momentum scale are typically below 0.1\% (0.4\%). The best measured momenta are obtained via the combination of the individual measurements from the ID and MS with scale uncertainties of about 0.04\% in the barrel region and up to about 0.2\% at $|\eta| > 2$. The results are checked by comparing the corrections extracted in both samples and also with an additional sample of 5 million \Upsmumu\ decays. These studies demonstrate the validity of the corrections and uncertainties in the range $6 < \pt \lesssim 100$~GeV.

\section{\Hgg}

\ \ \ \ The \Hgg\ decay offers a clean signature with a narrow peak from two energetic and isolated photons lying on top of a smoothly falling background which can be determined directly from data. The typical mass resolution is about 1.7~GeV and the signal-to-background ratio under the peak is $\sim 3\%$.

The analysis is based on a sample of two photon events with a purity of 75\%, selected by requiring tight identification criteria based on calorimeter shower shapes, isolation requirements and cuts on the ratio of the transverse momenta of the photons to the diphoton invariant mass ($\pt/\mgg$). The mass is calculated using the energies measured in the calorimeter and the opening angle between the photons, connecting the  impact points of the EM showers with the diphoton production vertex. The vertex is chosen by an algorithm that exploits the longitudinal segmentation of the calorimeter and information from the tracks associated to each vertex. The mass resolution is completely dominated by the resolution on the energy measurement.

In order to improve the Higgs boson mass determination, the events in each of the two datasets (7~TeV and 8~TeV) are split into ten categories with different signal-to-background ratios ($s/b$ from 2\% to 30\%), diphoton mass resolutions (1.2~GeV to 2.4~GeV) and systematic uncertainties (0.23\% to 0.59\%). The categories are defined according to the presence of converted photons, the $\eta$ of the photons and the component of the diphoton transverse momentum orthogonal to the diphoton thrust axis in the transverse plane (\ptt). The Higgs boson mass and signal strength ($\mu = \sigma/\sigma_{\mathrm{SM}}$) are obtained from a simultaneous signal-plus-background fit to the invariant mass spectrum in each category where the parameters associated to the background model are allowed to vary. Figure~\ref{fig:Hgg_spectrum} shows the sum across the categories, weighted by $s/b$ in each category, of the data points and the the analytical functions that model the signal and the background. The functions are chosen to minimise the statistical uncertainty and the expected bias on $\mu$ on simulated data. Systematic uncertainties are treated as nuisance parameters added to the model and constrained by Gaussian or Log-normal functions.

The systematic uncertainty on the determination of \mH\ is dominated by the uncertainties on the photon energy scale (ES): 0.17\% to 0.57\% depending on the category. The main contributions arise from non-linearities and quantities that affect differently the ES of electrons and photons, such as the relative calibration of the different gains used in the calorimeter readout, layer inter-calibration and mis-modellings of the material in front of the calorimeter. The maximum estimated bias on \mH\ from the choice of background modelling functions varies between 0.05\% and 0.2\%. The uncertainties on the diphoton mass resolution have a negligible impact on the mass measurement but a mild impact on the signal strength determination. The later is also affected by uncertainties on the the Higgs boson production cross section and branching ratio, luminosity, photon identification efficiency and other small contributions.

The measured mass in this channel is $\mH = 125.98 \pm 0.42 \mbox{ (stat)} \pm 0.28 \mbox{ (syst)}$~GeV = $125.98 \pm 0.50$~GeV, to be compared with the previous result~\cite{ATLAS_Higgs_diboson}: $\mH = 126.8 \pm 0.2 \mbox{ (stat)} \pm 0.7 \mbox{ (syst)}$~GeV. The change in the central value ($-0.8$~GeV) is consistent within one standard-deviation ($\sigma$) from the expected change due to the new calibration procedure ($-0.45$~GeV) and the induced fluctuations on \mgg\ ($\pm 0.35$~GeV) from fluctuations in the measured masses of individual events. The systematic uncertainty on \mH\ has been reduced by a factor 2.5. The statistical uncertainty is compatible with the expected value of 0.35~GeV for the measured signal strength $\mu = 1.29 \pm 0.30$ at the best-fit mass, and with the expectation for a standard model (SM) Higgs boson signal: 0.45~GeV for $\mu = 1$. This uncertainty increased compared to the previous result mainly due to the reduction of the measured signal strength. The most precise determination of $\mu$ from this data are based on an analysis optimised to measure the signal strength~\cite{ATLAS_Hgg_couplings}.

Several cross-checks have been performed to ensure the robustness of the mass determination. The combined mass has been compared to the measurement in different sets of the diphoton events, divided according to the number of converted photons, the number of primary vertices or the number of photons in the barrel and endcaps. No deviation above 1.5$\sigma$ is observed. The comparisons are shown in Fig.~\ref{fig:Hgg_deltaM}.

\begin{figure}[htb]
\centering
\subfloat[][\label{fig:Hgg_spectrum}]{\includegraphics[height=0.25\textheight]{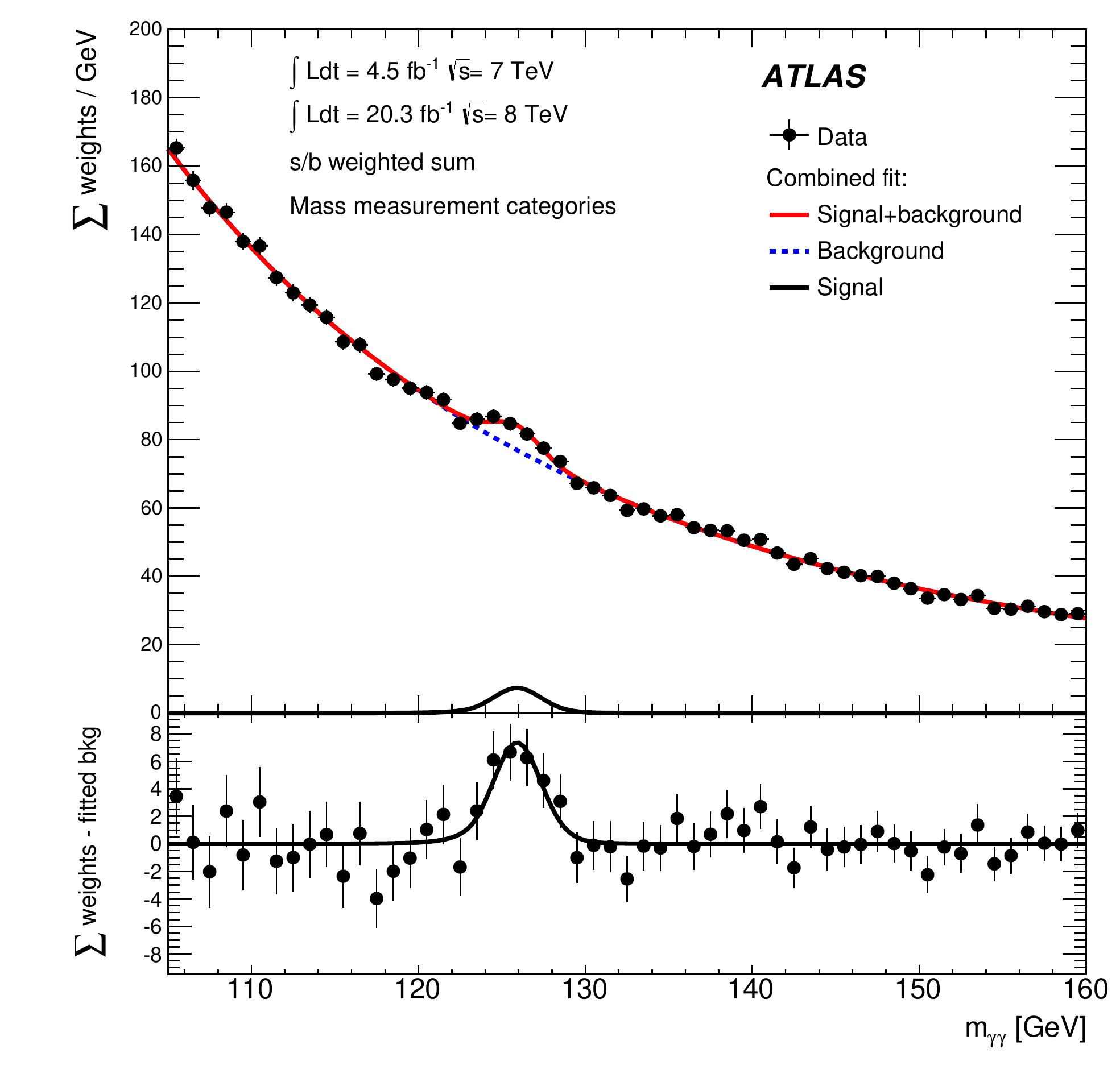}}
\subfloat[][\label{fig:Hgg_deltaM}]{\includegraphics[height=0.22\textheight]{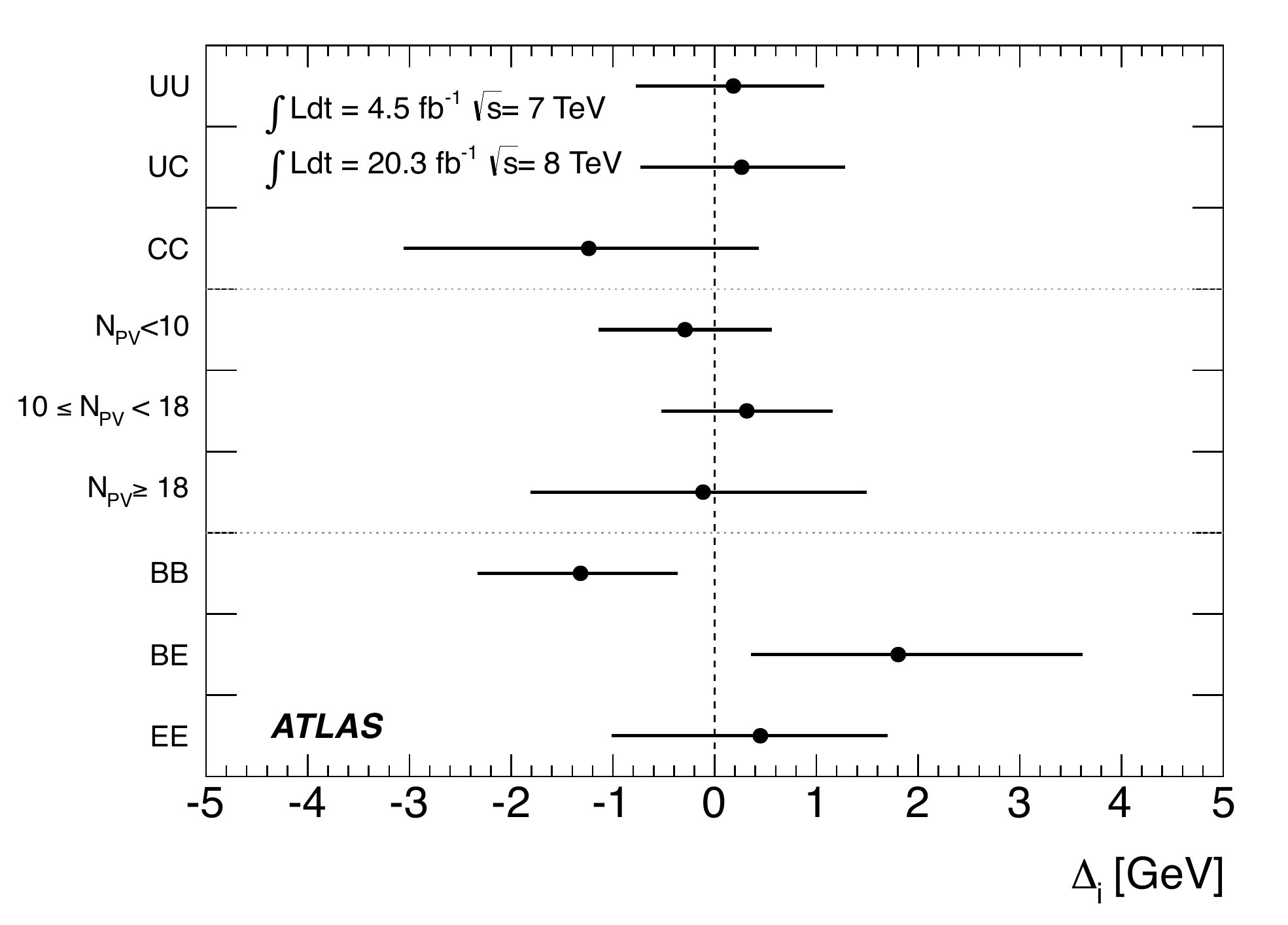}}
\caption{(a) Diphoton invariant mass spectrum with the sum of data points and both signal ($s$), background ($b$) and $s+b$ models weighted by $s/b$ in each category~\cite{ATLAS_Higgs_mass}. The bottom plot shows the difference between the summed weights and the background component of the fit. (b) Difference between the mass measured in a given subset of the diphoton events (see text) and the combined mass~\cite{ATLAS_Higgs_mass}.}
\end{figure}

\section{\H4l}

\ \ \ \ The \H4l\ decay channel has large high signal-to-background ratio $s/b \sim 1.6$ and high resolution on the mass determination, varying from 1.6~GeV to 2~GeV depending on the final state ($4\mu$, $2e2\mu$, $2\mu2e$ and $4e$), although low event yields. The main background to this channel is the irreducible decay of pairs of $Z$ bosons into electrons and muons (\ZZ4l), with small contributions from the reducible $Z$ + jets and top-quark pair production (\ttbar) processes.

Higgs boson candidates are selected by requiring two pairs of same-flavour, oposite-sign leptons. Each lepton must satisfy identification quality criteria and requirements on the impact parameter with respect to the primary collision vertex and isolation. The four-lepton invariant mass (\m4l) resolution is improved by 15\% with the addition of at most one final-state radiation (FSR) photon and the application of a kinematic fit to constrain the mass of the leading lepton pair to the $Z$ pole mass within the experimental resolution. After applying cuts on the masses of both lepton pairs, the total Higgs boson signal selection efficiency varies from 39\% to 20\% from the $4\mu$ to the $4e$ final states at $\mH = 125$~GeV.

The \ZZ4l\ background is estimated from simulation while the yields of the reducible background are normalised from control regions by relaxing identification, isolation or impact parameter requirements. A boosted decision tree ($BDT_{ZZ^*}$) trained on simulated signal ($\mH = 125$~GeV) and background events is applied to reduce the impact of the \ZZ4l\ background. It uses the transverse momentum and the pseudorapidity of the four-lepton system, plus a matrix-element-based kinematic discriminant. The signal is extracted with a two-dimensional fit to \m4l\ and the $BDT_{ZZ^*}$ output, which brings 8\% extra sensitivity on the mass with respect to a simple fit to \m4l. The invariant mass distribution of the selected events is shown in Fig.~\ref{fig:H4l_spectrum}, while the data points in the \m4l\ versus $BDT_{ZZ^*}$ plane are shown in Fig.~\ref{fig:H4l_2D}. Superimposed are the models of the signal and \ZZ4l\ background, both taken from simulation, together with the model of the reducible background, determined via data-driven techniques.


The measured Higgs boson mass in the \H4l\ channel is $\mH = 124.51 \pm 0.52 \mbox{ (stat)} \pm 0.06 \mbox{ (syst)}$~GeV = $124.51 \pm 0.52$~GeV. The systematic uncertainty is obtained from the quadrature subtraction of the fit uncertainty evaluated with and without the systematic uncertainties fixed at their best fit values, with a precision around 10~MeV. The present result is compatible among all final states and also with the previous result $\mH = 124.51^{+0.6}_{-0.5} \mbox{ (stat)} ^{+0.5}_{-0.3} \mbox{ (syst)}$~GeV. The measured signal strength is $\mu = 1.66 \pm 0.45$, consistent with the SM expectation. The most precise results for $\mu$ from this data are based on an analysis described in Ref.~\cite{ATLAS_H4l_couplings}.

\begin{figure}[htb]
\centering
\subfloat[][\label{fig:H4l_spectrum}]{\includegraphics[height=0.22\textheight]{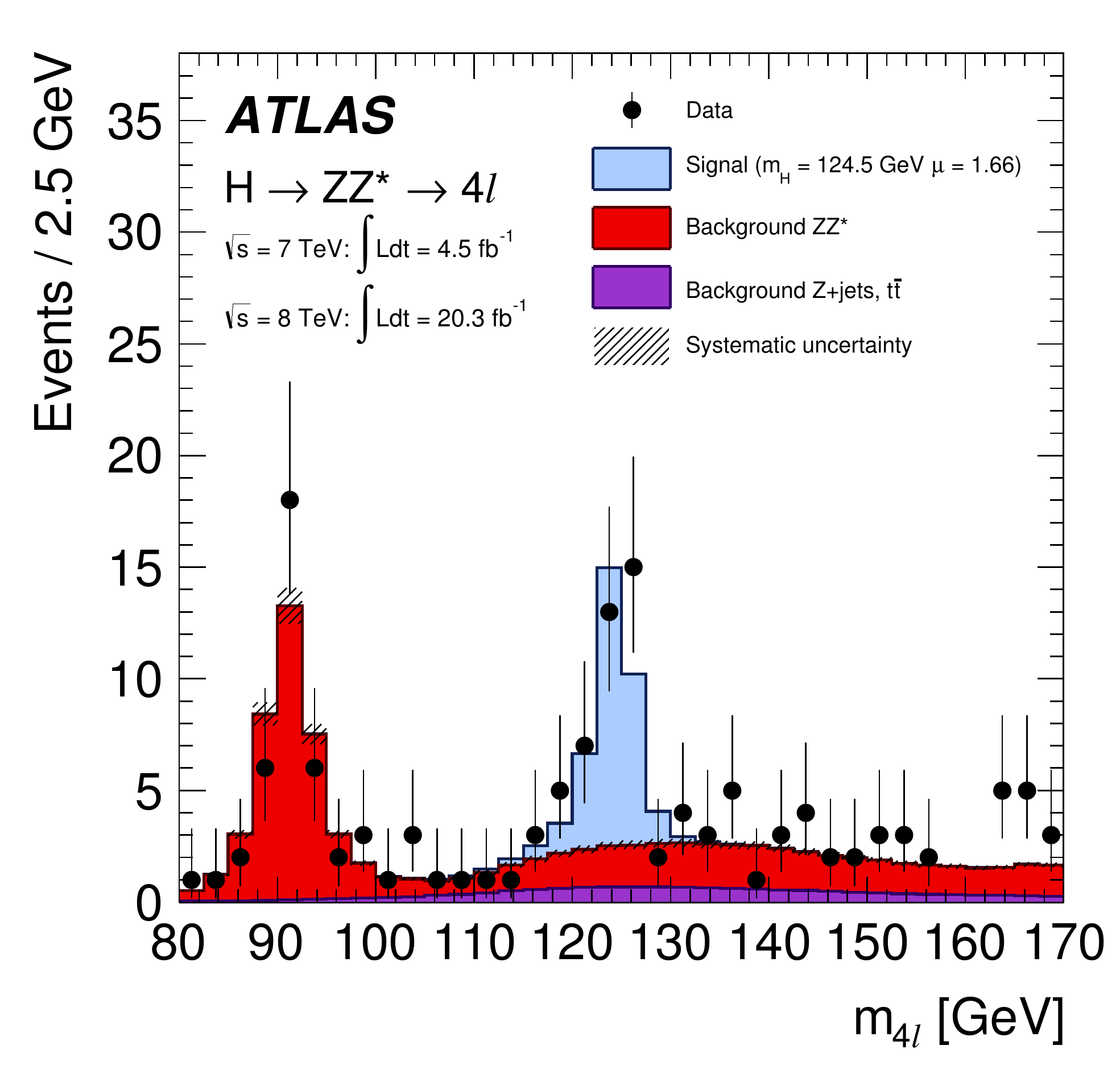}}
\subfloat[][\label{fig:H4l_2D}]{\includegraphics[height=0.22\textheight]{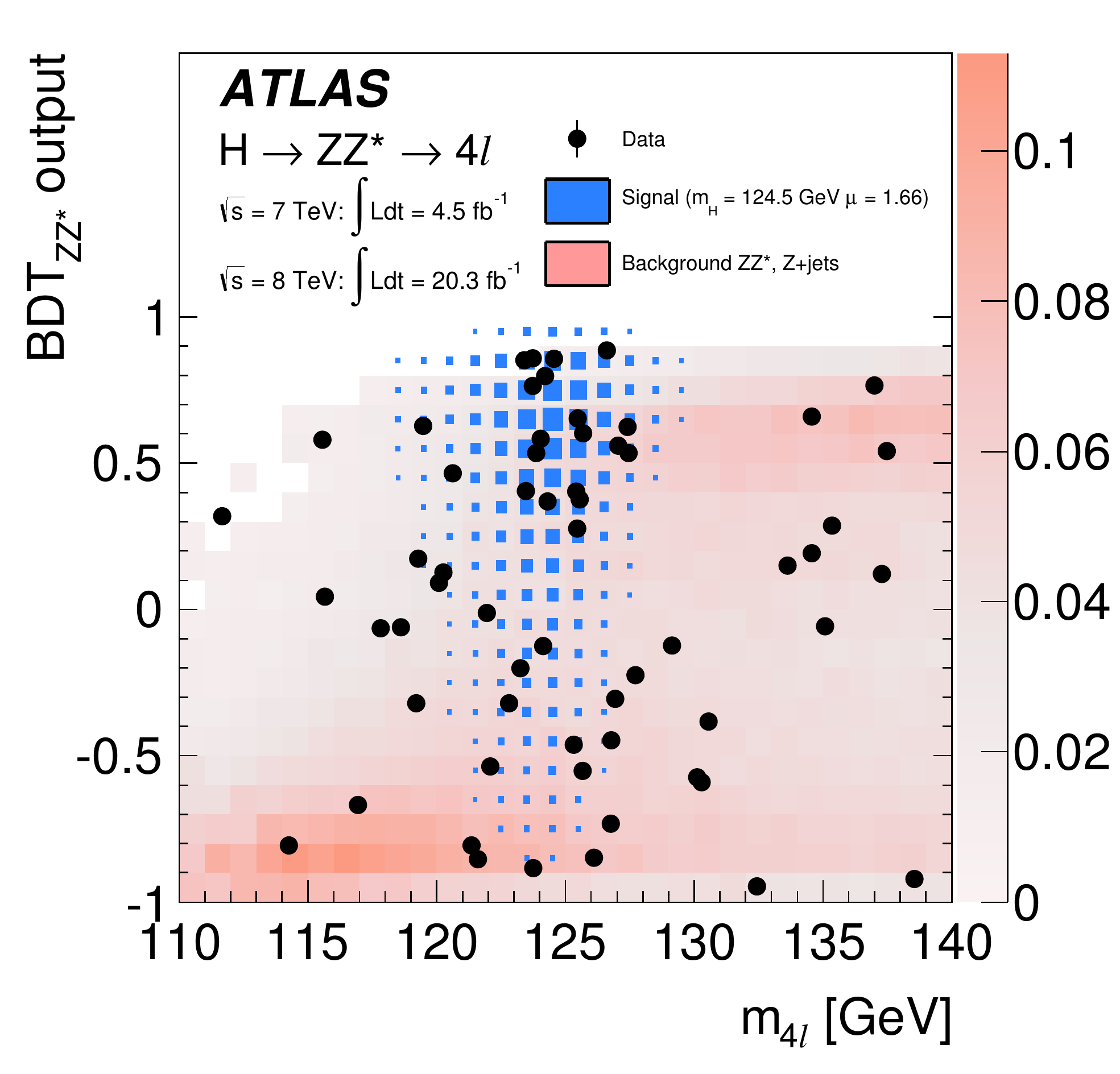}}
\caption{Four-lepton invariant mass (\m4l) distribution (a) and \m4l\ versus $BDT_{ZZ^*}$ output in the \m4l\ range $110-140$~GeV (b) for the selected events~\cite{ATLAS_Higgs_mass}. The expected distributions for the signal at $\mH = 124.5$~GeV normalised by the measured signal strength and backgrounds are superimposed.}
\end{figure}

\section{\HWW}

\ \ \ \ The \HWW\ decay has a relatively large rate and $s/b$ of $\mathcal{O}(10)\%$ but limited mass resolution due to the presence of neutrinos in the final state. The dominant backgrounds to this channel are processes containing two $W$ bosons in the final state ($WW$, \ttbar\ and single-top production), with additional contributions from other diboson processes, $W$ + jets with jets faking one of the leptons and $Z/\gamma^*$.

The analysis, described in Ref.~\cite{ATLAS_Higgs_diboson}, splits events with two opposite-sign leptons and significant missing transverse energy into three bins according to the jet multiplicity. Each bin has different contributions from the various background processes and signal production modes -- mainly gluon-fusion in 0- and 1-jet bins and a significant contribution of vector-boson fusion in the 2-jets bin. The backgrounds are normalised from control-regions in data ($WW$ and top-quark production), modelled using data driven techniques ($W$+jets and $Z/\gamma^*$) or taken directly from simulated events (other diboson processes). Figure~\ref{fig:HWW_0+1jets} shows the transverse mass distributions in the 0- and 1-jet bins for all lepton flavours, while Fig.~\ref{fig:HWW_2jets} shows the same quantity in the 2-jets bin only in the $e/\mu$ final state. An excess of events from the background-only expectation is observed with a significance of $3.8\sigma$ at $\mH = 125.5$~GeV, corresponding to $\mu = 1.00^{+0.32}_{-0.29}$.

\begin{figure}[htb]
\centering
\subfloat[][\label{fig:HWW_0+1jets}]{\includegraphics[height=0.22\textheight]{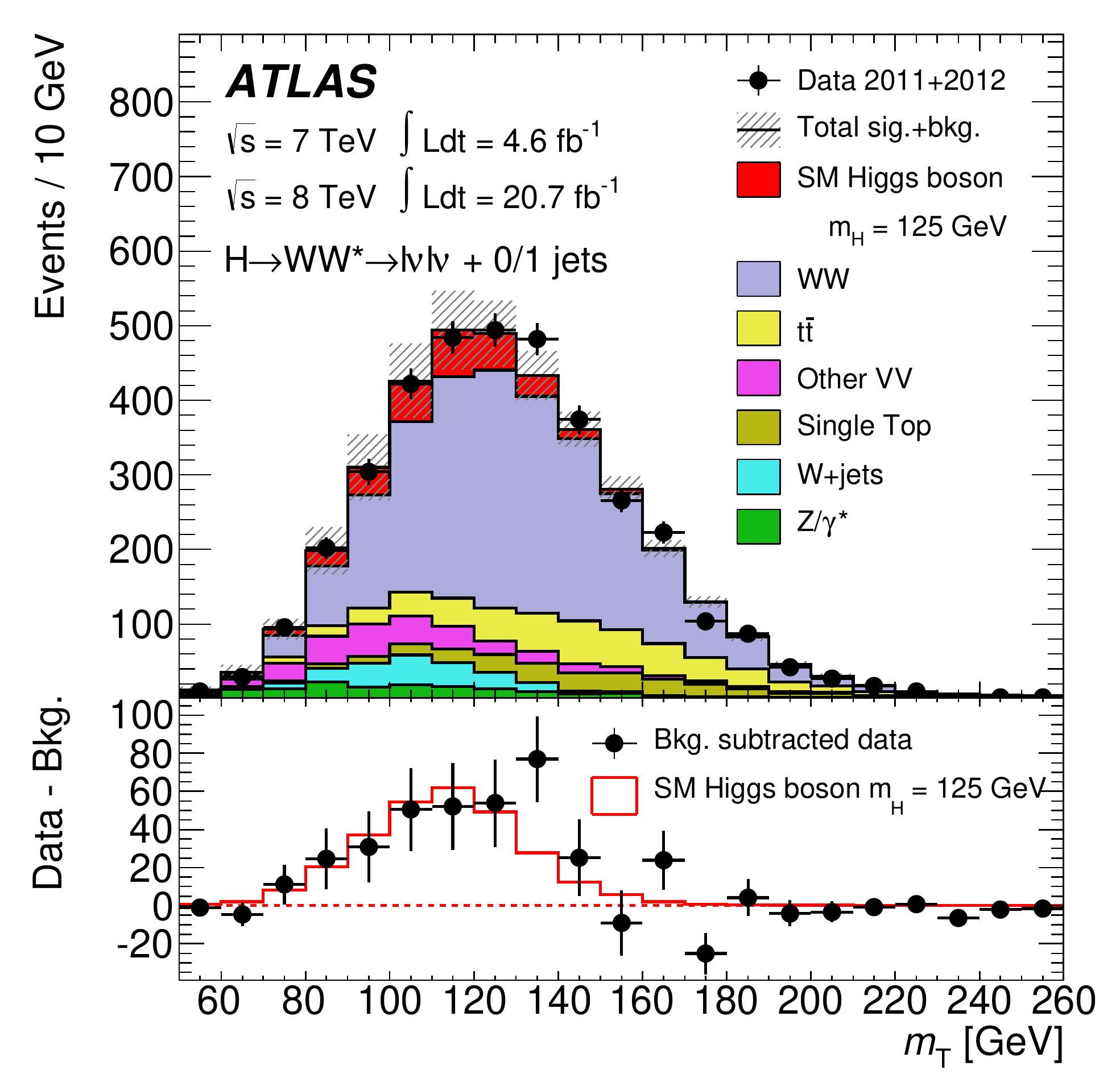}}
\subfloat[][\label{fig:HWW_2jets}]{\includegraphics[height=0.22\textheight]{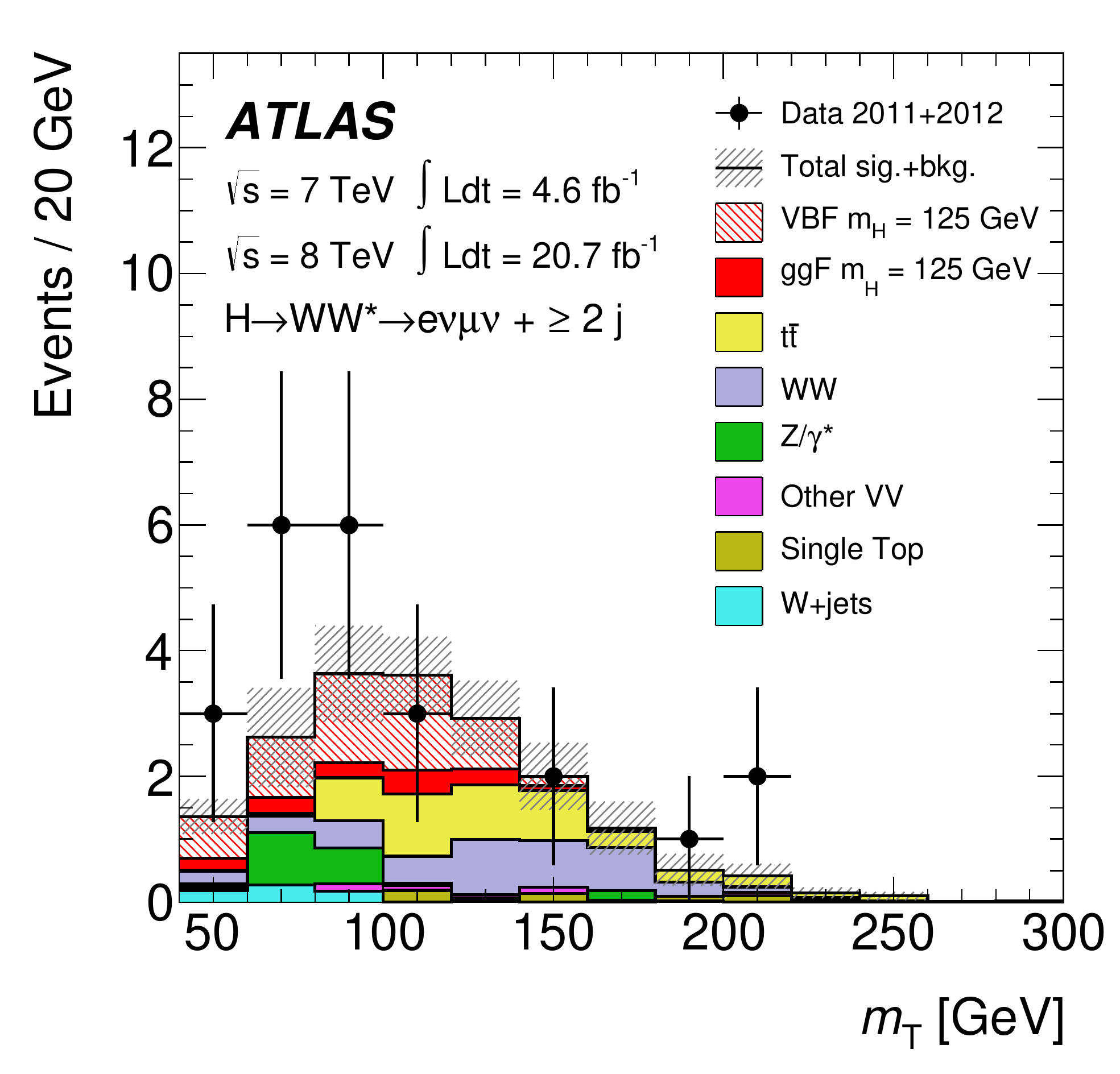}}
\caption{Transverse mass (\mT) distributions in the 0/1-jet (a) and 2-jets (b) bins for the \HWW\ analysis~\cite{ATLAS_Higgs_diboson}. The hatch area represents the uncertainties on both signal and background yields. The bottom inset on (a) shows the background-subtracted distribution compared with the SM Higgs boson signal.}
\label{fig:HWW}
\end{figure}

\section{\HZg}
\vspace{-1mm}

\ \ \ \ The observation of the \HZg\ signal, with \Zll, is challenging given the low expected yields and the large backgrounds, resulting in $s/b$ of $\mathcal{O}(1\%)$. The dominant background is the $Z+\gamma$ process, accounting for 82\% of the events after the event selection described in Ref.~\cite{ATLAS_HZgamma}, which requires two leptons of opposite-sign with a mass consistent with the $Z$ pole mass and a photon satisfying tight identification and isolation criteria.

The presence of a signal is tested via a fit to the dilepton-plus-photon invariant mass distribution (\mllg), shown in Fig.~\ref{fig:HZg_spectrum}, where the signal and the background are modelled by analytical functions independently in eight event categories. The categories are defined according to the centre-of-mass energies, the lepton flavours and additionally $\Delta\eta(Z\gamma)$ and \ptt, which improve the sensitivity by 30\%. No significant deviations from the background only hypothesis are observed and limits are set on the ratio of the observed cross section to the SM prediction for the Higgs boson signal, as shown in Fig.~\ref{fig:HZg_limit}. The 95\% confidence level limit at $\mH = 125.5$~GeV is 11 times the SM prediction, compared an expectation of 9~$\times$~SM.

\vspace{-1mm}
\begin{figure}[htb]
\centering
\subfloat[][\label{fig:HZg_spectrum}]{\includegraphics[height=0.2\textheight]{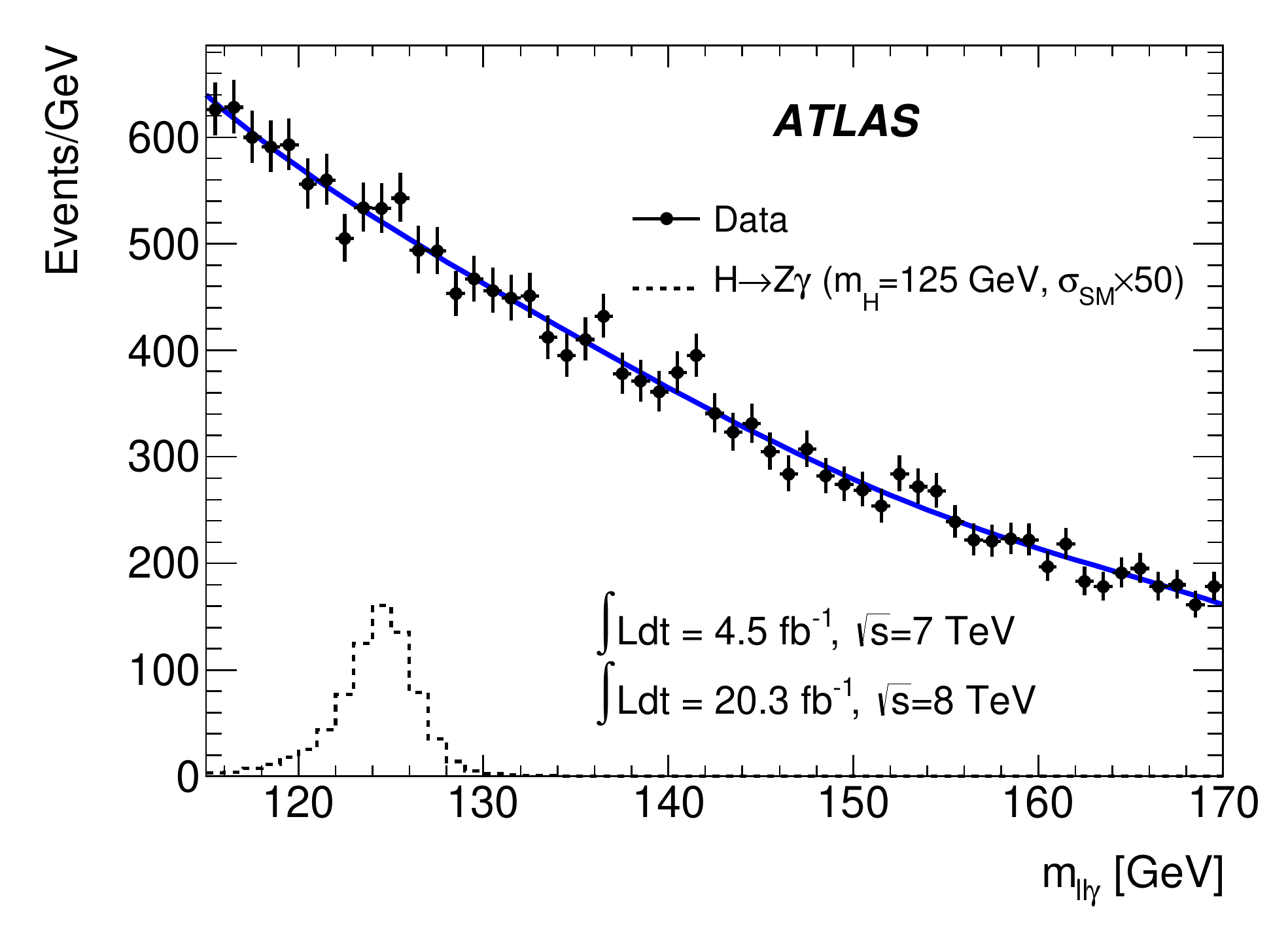}}
\subfloat[][\label{fig:HZg_limit}]{\includegraphics[height=0.2\textheight]{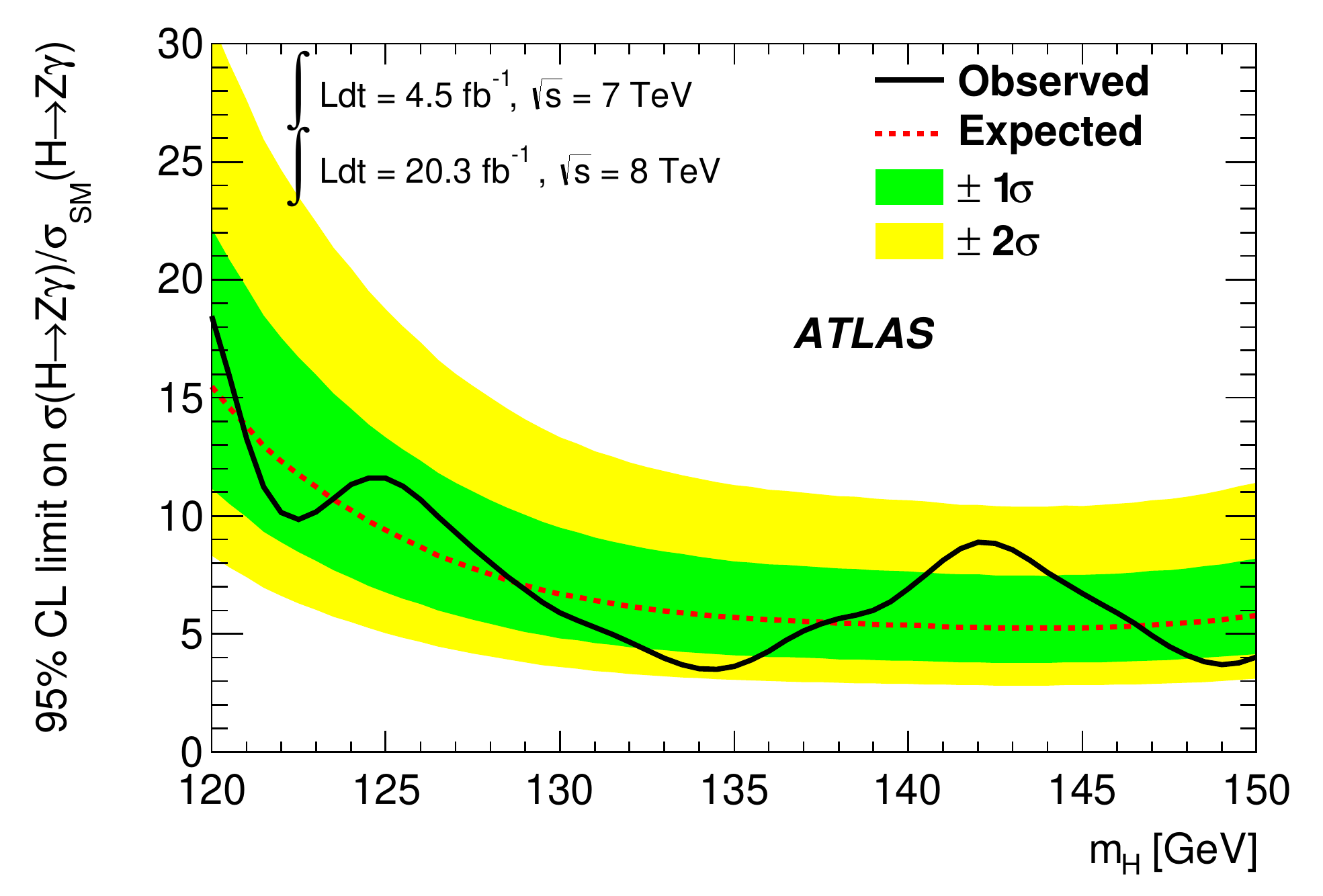}}
\vspace{-0.5mm}
\caption{(a) Distribution of \mllg\ superimposed with the background model and the expected SM signal at $\mH = 125$~GeV scaled by a factor 50~\cite{ATLAS_HZgamma}. (b) Limits on the production cross section of a SM Higgs boson decaying to $Z\gamma$ normalised to the SM expectation~\cite{ATLAS_HZgamma}.}
\end{figure}

\vspace{-3mm}
\section{Conclusions}
 
\ \ \ \ The analyses of the Higgs boson decays to bosonic channels using 25~\ifb\ of $pp$ collision data collected by the ATLAS experiment have been presented. An improved measurement of the Higgs boson mass is obtained from the results from the decay channels \Hgg\ and \H4l, which are compatible within 2.0 standard deviations. This result $\mH = 125.36 \pm 0.37 \mbox{ (stat)} \pm 0.18 \mbox{ (syst)} \mbox{ GeV}$ $= 125.36 \pm 0.41$~GeV supersedes the previous result from ATLAS.

\end{document}